\documentclass[twocolumn,english,aps,pra,superscriptaddress,showkeys,tightenlines]{revtex4}
\usepackage{amsmath}
\usepackage{amssymb}
\usepackage{graphicx}
\usepackage{color}

\begin{document}

\title{Spatial dependent spontaneous emission of an atom in a semi-infinite
waveguide of rectangular cross section}
\author{Hai-Xi Song}
\affiliation{Key Laboratory of Low-Dimensional Quantum Structures and Quantum Control of
Ministry of Education, Department of Physics and Synergetic Innovation
Center of Quantum Effects and Applications, Hunan Normal University,
Changsha 410081, China}
\author{Xiao-Qi Sun}
\affiliation{Key Laboratory of Low-Dimensional Quantum Structures and Quantum Control of
Ministry of Education, Department of Physics and Synergetic Innovation
Center of Quantum Effects and Applications, Hunan Normal University,
Changsha 410081, China}
\author{Jing Lu}
\affiliation{Key Laboratory of Low-Dimensional Quantum Structures and Quantum Control of
Ministry of Education, Department of Physics and Synergetic Innovation
Center of Quantum Effects and Applications, Hunan Normal University,
Changsha 410081, China}
\author{Lan Zhou}
\thanks{Corresponding author. Fax No.: +86-731-8887-3055}
\email{zhoulan@hunnu.edu.cn}
\affiliation{Key Laboratory of Low-Dimensional Quantum Structures and Quantum Control of
Ministry of Education, Department of Physics and Synergetic Innovation
Center of Quantum Effects and Applications, Hunan Normal University,
Changsha 410081, China}
\date{\today }

\begin{abstract}
We study a quantum electrodynamics (QED) system made of an two-level atom
and a semi-infinite rectangular waveguide, which behaves as a perfect mirror
in one end. The spatial dependence of the atomic spontaneous emission has
been included in the coupling strength relevant to the eigenmodes of the
waveguide. The role of retardation is studied for the atomic transition
frequency far away from the cutoff frequencies. The atom-mirror distance
introduces different phases and retardation times into the dynamics of the
atom interacting resonantly with the corresponding transverse modes. It is
found that the upper state population decreases from its initial as long as
the atom-mirror distance does not vanish, and is lowered and lowered when
more and more transverse modes are resonant with the atom. The atomic
spontaneous emission can be either suppressed or enhanced by adjusting the
atomic location for short retardation time. There are partial revivals and
collapses due to the photon reabsorbed and re-emitted by the atom for long
retardation time.
\end{abstract}

\maketitle


\narrowtext


\section{Introduction}

Any quantum system inevitably interacts with its surroundings which possess
a huge amount of uncontrollable degrees of freedom. Such interaction causes
the rapid destruction of quantum coherence, which is an essential
requirement for quantum information processing to fully exploit the new
possibilities opened by quantum mechanics. For example, the information
stored in two-level systems (we refer to atoms hereafter) can be destroyed
by their surrounding electromagnetic field. Spontaneous emission (SE) is one
of the most prominent effects in the interaction of atoms with vacuum. It is
an atomic radiation as follow: an atom initially in an excited state relax
to its ground state and emit a quanta of energy to its surrounding vacuum
electromagnetic (EM) field, which carries away the difference in energy
between the two levels.

SE is not only useless but also harmful to quantum information process.
However, recently studies have shown that it is useful to build a device in
a quantum network for controlling single photons by a local atom, e.g. the
atomic radiation leads to the total reflection of the single-photon
propagating in one quantum channel~\cite{Fans,ZLPRL08}, the frequency
converter for single photons~\cite{ZLfconvert}, and the transfer of single
photons from one quantum channel to the other~\cite{ZLQrouter}. Nowadays,
great interest has been paid on the use of atoms to act as a quantum node in
extended communication networks and scalable computational devices~\cite%
{zlZeno,Zheng,Roy,ZLMZI,GongPRA78,LawPRA78,SPT13,ZLQrouter2,PRL116,attenuator,MTcheng,Longo,Alexanian,TShiSun}%
. As the SE rate of a single atom can be modified by a succession of short
and strong pulses or measure to the quantum system, a dynamical
quantum-Zeno-effect(QZE) switch for single photons is proposed~\cite{zlZeno}%
. The quantum interference between multiple transition pathways of atomic
internal states has been exploited to modify the transport property of the
single photons in a quantum channel~\cite{GongPRA78}. With the well-known
result that the atomic SE depends on the electromagnetic vacuum environment
that the atom is subjected to, a boundary has been used to increase the
efficiency of the quantum router~\cite{ZLQrouter2}.

Actually, the SE rate of a single atom is one of the basic topic of quantum
electrodynamics, numerous studies of the SE rates~\cite{PR219-263} have been
carried out for atoms in free space~(e.g.\cite{PRA50-1755}), in a
cavity~(e.g. \cite{PRL47-233}), near a metallic mirror~(e.g. \cite%
{PRA35-5081}) or a dielectric interfacer~(e.g. \cite{PRD3-280}), between two
mirrors~(e.g. \cite{PRA43-5795}) or two dielectric interfaces~(e.g. \cite%
{PRA57-3913}). However, photons used to transmit information or distribute
the entanglement along the network, are confined in a one dimensional (1D)
waveguide. Similar to cavities, 1D waveguide has a well-defined mode
spectrum and a relatively loss-free environment. However, unlike cavities,
modes are available in waveguide for photons to propagate. The geometry
constraint not only confines the propagating direction of photon, but also
gives rise to an increasing of the interference effects. Since photons do
not interact with each other, atoms implanted in waveguide are necessary to
mediate the photon-photon interaction or redirect the possible propagating
directions. The coupling strength of atoms to the waveguide is enhanced by
decreasing the mode volume. Consequently, the atomic radiation in 1D
waveguide plays an important role in controlling photons in quantum
networks. And the studies on the atoms in 1D waveguide is now referred to by
the term ``waveguide quantum electrodynamics (QED).''

Since the radiative properties of an atom in a confined space differ
fundamentally from that in free space, considerable interest has been paid
on atoms in a semi-infinite or infinite 1D waveguide. However most works
focus on 1D waveguide without a cross section~\cite%
{DongPRA79,NJP14-043,PRA87-013820,PRA87-063830,PRA91-053845}. In this paper,
we study the radiative properties of an atom in a semi-infinite waveguide of
rectangular cross section, which is a typical 1D QED system. The termination
of the waveguide is regarded as a perfect mirror, which reflects emitted
photons back to the atom. We analyze the interaction of an initially excited
two-level atom with the waveguide in vacuum. The Markovian approximation is
first used to analyze the dependence of the SE rate on the density of states
and the spatial profile of the waveguide. To find the influence of the time
that one-photon wave packet requires to bounce back and forth between the
atom and the mirror, we perform the linear expansion of the dispersion
relation for the atomic transition frequency far away from the cutoff
frequencies, and obtain a delay-differential equation. Then the atomic SE
dynamics is studied by varying the cross section of the waveguide as well as
the atomic location.

This paper is organized as follows. In Sec.~\ref{Sec:2}, we introduce the
system we studied. In Sec.~\ref{Sec:3}, we derive the relevant equations
describing the dynamics of the system in single-excitation subspace. In Sec.~%
\ref{Sec:4}, we do the Markovian approximation to study the effect of the
mode profile on the spontaneous rate. In Sec.~\ref{Sec:5}, the atomic
dynamics is studies by linearly expanding the dispersion relation around the
transition frequency which is valid far from the branch threshold, where the
delay time is introduced. Finally, We conclude this work in Sec.~\ref{Sec:5}.


\section{\label{Sec:2}An atom inside a rectangular pipe waveguide}

\begin{figure}[tbp]
\includegraphics[bb=48bp 523bp 550bp 751bp,width=8cm]{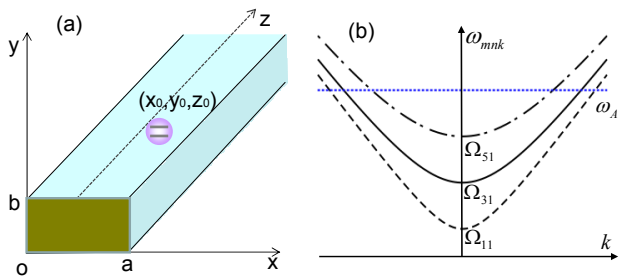}
\caption{(Color online) Schematic illustration for (a) the two-level atom
embedded in a semi-infinite waveguide of rectangular cross section, and (b)
the dispersion relation of the guiding modes which interact with the atom at
$\vec{r}=\left( a/2,b/2,z_{0}\right) $ with $a=2b$.}
\label{fig:1}
\end{figure}

The system we studied is shown in Fig.~\ref{fig:1}. A waveguide made of
ideal perfect conducting walls is formed from surfaces at $x=0$, $x=a$, $y=0$%
, $y=b$, $z=0$, and is placed along the $z$ axis.
The waveguide is assumed to be infinite along the $z$ direction. The
boundary condition restricts photons to travel without loss of power in two
independent guiding modes \cite{liqongPRA,HuangPRA}: TE modes whose electric
field has no longitudinal component, and TM modes whose magnetic field has
no longitudinal component. Let $\vec{k}=\left( k_{x},k_{y},k\right) $ be the
wave vector. The relations $k_{x}=m\pi /a$ and $k_{y}=n\pi /a$ with positive
integers $n,m\ $can be imposed by the condition that the tangential
components of the electric field vanish at all the conducting wall, however,
there is no constraint on $k$. Therefore, the waveguide allows a continuous
range of frequencies described by the dispersion relation%
\begin{equation}
\omega _{mnk}=\sqrt{c^{2}k^{2}+\Omega _{mn}^{2}},  \label{A-1}
\end{equation}%
where $c$ is the speed of light in vacuum, $\Omega _{mn}=\pi c\sqrt{%
m^{2}/a^{2}+n^{2}/b^{2}}$ is the minimum frequency for a traveling wave. We
note that $m$ and $n$ cannot both be zero. If $a>b$, TE$_{10}$ is the lowest
guiding mode for the waveguide \cite{EMtextb}, and the lowest TM modes occur
for $m=1$, $n=1$. Obviously, the waveguide modes form a one-dimensional
continuum. Each guiding mode provides a quantum channel for photons to
travel from one location to the other.

At $\vec{r}=\left( x_{0},y_{0},z_{0}\right) $ is an atom with transition
frequency $\omega _{A}$ between upper level $\left\vert e\right\rangle $ and
lower level $\left\vert g\right\rangle $, which is excited initially. The
atom sits inside the hollow waveguide. $z_{0}$ is the distance between the
atom to the wall (or the mirror) at $z=0$. The free Hamiltonian for the atom
is described by
\begin{equation}
H_{s}=\omega _{A}\sigma _{+}\sigma _{-},  \label{A-2}
\end{equation}%
where $\sigma _{+}\equiv \left\vert e\right\rangle \left\langle g\right\vert
$ ($\sigma _{-}\equiv \left\vert g\right\rangle \left\langle e\right\vert $)
is the rising (lowering) atomic operator. For the purpose of simplicity, the
electric dipole of the stationary atom is assumed to be oriented along the $z
$ direction, which means that the atom only interacts with the TM$_{mn}$
modes. Since the number $\left( m,n,k\right) $ specifies the mode function
of this air-filled metal pipe waveguide, we label the annihilation operator
for each TM guiding mode by $a_{mnk}$. The free Hamiltonian for the
waveguide is described by%
\begin{equation}
H_{f}=\sum_{mn}\int_{-\infty }^{\infty }dk\text{ }\omega _{mnk}a_{mnk}^{\dag
}a_{mnk}  \label{A-3}
\end{equation}%
The interaction between the atom and field via the dipole coupling in the
rotating-wave approximation reads%
\begin{equation}
H_{I}=\sum_{mn}\int_{-\infty }^{\infty }dkg_{mnk}(\sigma _{-}a_{mnk}^{\dag
}-\sigma _{+}a_{mnk})  \label{A-4}
\end{equation}%
where the coupling strength%
\begin{equation}
g_{mnk}=\frac{2id\Omega _{mn}}{\sqrt{\pi \epsilon _{0}A\omega _{mnk}}}\sin
\frac{m\pi x_{0}}{a}\sin \frac{n\pi y_{0}}{b}\cos (kz_{0})  \label{A-5}
\end{equation}%
Here, $\epsilon _{0}$ the permittivity of free space, $d$ the magnitude of
the transition dipole moment of the atom and assumed to be real, $A=ab$ the
area of the rectangular cross section. Using the dispersion relation, the
coupling strength can be rewritten as%
\begin{eqnarray}
g_{mn\omega } &=&\frac{2id\Omega _{mn}}{\sqrt{\pi \epsilon _{0}A\omega }}%
\sin \frac{m\pi x_{0}}{a}\sin \frac{n\pi y_{0}}{b}  \label{A-6} \\
&&\times \cos \left( \sqrt{\omega ^{2}-\Omega _{mn}^{2}}\frac{z_{0}}{c}%
\right) .  \notag
\end{eqnarray}%
The cosine function in Eqs. (\ref{A-5}) and (\ref{A-6}) occurs due to the
termination of the waveguide, which presents the difference from the
infinite waveguide. Obviously, the atom located at $x_{0}=a/2$ and $y_{0}=b/2
$ decouples to the TM$_{mn}$ guiding mode with even integer $m$ or $n$. The
total system are described by Hamiltonian $H=H_{s}+H_{f}+H_{I}$.


\section{\label{Sec:3}Evolution of the atom-vacuum system}


In this section we study the dynamics of the system when the atom is
initially in the excited state $\left\vert e\right\rangle $ and the field is
in the vacuum state $\left\vert 0\right\rangle $. Since the number of quanta
is conserved in this system, we can write the wavefunction of the system as:%
\begin{equation}
\left\vert \psi \left( t\right) \right\rangle =\varepsilon \left( t\right)
\left\vert e0\right\rangle +\sum_{mn}\int dk\varphi _{mn}\left( k,t\right)
a_{mnk}^{\dag }\left\vert g0\right\rangle  \label{B-1}
\end{equation}%
in one quantum subspace. The first term in Eq.(\ref{B-1}) describes the atom
in the excited state with no photons in the field, $\varepsilon \left(
t\right) $ is the corresponding amplitude, whereas the second term in Eq.(%
\ref{B-1}) describes the atom in the ground state with a photon emitted at a
mode $k$ of the TM$_{mn}$ guiding mode, $\varphi _{mn}\left( k,t\right) $ is
the corresponding amplitude. The initial state of the system is denoted by
the amplitudes $\varepsilon \left( 0\right) =1$, $\varphi _{mn}\left(
k,0\right) =0$. The Schr\"{o}dinger equation results in the following
coupled equation of the amplitudes
\begin{subequations}
\label{B-2}
\begin{eqnarray}
\dot{\varepsilon}\left( t\right) &=&-i\omega _{A}\varepsilon \left( t\right)
+i\sum_{mn}\int_{-\infty }^{\infty }dkg_{mnk}\varphi _{mn} , \\
\dot{\varphi}_{mn}&=&-i\omega _{mnk}\varphi _{mn}\left( k,t\right)
-ig_{mnk}\varepsilon (t),
\end{eqnarray}%
where the overdot indicates the derivative with respect to time. The
population of the atomic excited state are usually analyzed by eliminating
the field variables and focusing on the dynamics of the radiating system. We
start by removing the high frequency term in Eq.(\ref{B-2}) via the
transformation
\end{subequations}
\begin{equation}
\varepsilon \left( t\right) =\tilde{\varepsilon}\left( t\right) e^{-i\omega
_{A}t},\varphi _{mn}\left( k,t\right) =\tilde{\varphi}_{mn}\left( k,t\right)
e^{-i\omega _{mnk}t}.  \label{B-3}
\end{equation}%
Then we formally integrating equation of $\tilde{\varphi}_{mn}\left(
k,t\right) $, which is later inserted into the equation for $\tilde{%
\varepsilon}\left( t\right) $. The probability amplitude for the excited
atomic state is determined by the following integro-differential equation
\begin{equation}
\partial _{t}\tilde{\varepsilon}\left( t\right) =\sum_{mn}\int_{-\infty
}^{\infty }dk\int_{0}^{t}d\tau g_{mnk}^{2}\tilde{\varepsilon}\left( \tau
\right) e^{i\left( \omega _{A}-\omega _{mnk}\right) \left( t-\tau \right) }.
\label{B-4}
\end{equation}%
Eq.(\ref{B-4}) shows that the value of $\partial _{t}\tilde{\varepsilon}%
\left( t\right) $ depends on the values of $\tilde{\varepsilon}\left(
t\right) $ at all earlier time.


\section{\label{Sec:4}Spatial dependence of the spontaneous rate}

\begin{figure}[tbp]
\includegraphics[bb=42bp 589bp 539bp 759bp,width=8cm]{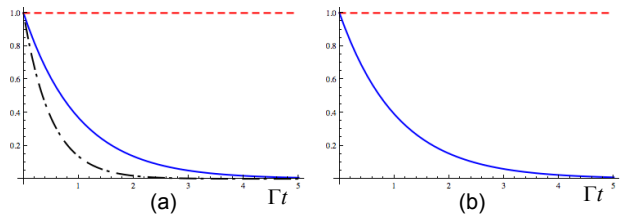}
\caption{(Color online) Atomic excitation probability as function of $\Gamma
t$ with $a=2b$. (a) The transition frequency $\protect\omega _{A}=(\Omega
_{11}+\Omega _{31})/2$. The atom is located at three different positions $%
z_{0}=0$ (black dot-dashed line), $z_{0}=\protect\lambda _{1A}/8$ (blue
solid line), $z_{0}=\protect\lambda _{1A}/4$ (red dashed line). (b) The
transition frequency $\protect\omega _{A}=(\Omega _{31}+\Omega _{51})/2$.
The atom is located at $z_{0}=\protect\lambda _{1A}/4$. The red dashed line
presents the contribution of the TM$_{11}$ mode. The blue solid line gives
the contribution of both TM$_{11}$ and TM$_{31}$ modes. }
\label{fig:2}
\end{figure}

To see how the mode distribution of the quantum vacuum fluctuation modifies
the atomic spontaneous rate, we set $\tilde{\varepsilon}\left( \tau \right)
\approx \tilde{\varepsilon}\left( 0\right) =1$ on the right-hand side of
equation (\ref{B-4}), and the amplitude of level $\left\vert e\right\rangle $
reads%
\begin{equation}
\varepsilon \left( t\right) =e^{-i\omega _{A}t}\left[ 1+\int_{0}^{t}d\tau
\left( t-\tau \right) G\left( \tau \right) e^{i\omega _{A}\tau }\right] ,
\label{B-5}
\end{equation}%
where the reservoir response (memory) function%
\begin{equation}
G\left( \tau \right) =-\int_{-\infty }^{\infty
}dk\sum_{mn}g_{mnk}^{2}e^{-i\omega _{mnk}\tau }  \label{B-6}
\end{equation}%
characterizes the spectrum of the rectangular hollow waveguide. Its Fourier
transformation yields the coupling spectrum~\cite{N405-546,PRL87-270}
\begin{equation}
G\left( \omega \right) =-\sum_{mn}g_{mn\omega }^{2}\rho \left( \omega \right)
\label{B-7}
\end{equation}%
which is the density of states
\begin{equation}
\rho \left( \omega \right) =\frac{\omega }{c\sqrt{\omega ^{2}-\Omega
_{mn}^{2}}}  \label{B-8}
\end{equation}%
weighted by the strength of the coupling to the continuum. Since the
dispersion relation of the semi-infinite waveguide is the same as that of
the infinite rectangular waveguide, $\rho \left( \omega \right) $ in Eq.(\ref%
{B-8}) is also the density of state of the infinite rectangular waveguide.
For weak atom-field coupling, the amplitude of level $\left\vert
e\right\rangle $ decays exponentially%
\begin{equation*}
\varepsilon \left( t\right) \approx \exp \left( -i\omega _{A}t-\frac{1}{2}%
Rt\right)
\end{equation*}%
Accordingly, the SE rate, the key ingredient in the SE dynamics, reads
\begin{equation}
R=2\pi \int_{-\infty }^{+\infty }d\omega f\left( \omega \right) G\left(
\omega \right)  \label{B-9}
\end{equation}%
which is the overlap of the coupling spectrum $G\left( \omega \right) $ and
the modulation spectrum~\cite{N405-546,PRL87-270}
\begin{equation}
f\left( \omega \right) =\frac{t}{2\pi }\text{sinc}^{2}\frac{\left( \omega
-\omega _{A}\right) t}{2}.  \label{B-10}
\end{equation}%
Here, $\sin c\left( x\right) =\sin x/x$. the modulation spectrum is
symmetrically centered on $\omega _{a}$ and decays in amplitude as $t^{-1}$.
Function $f\left( \omega \right) $ is the Fourier transform of the function
\begin{equation}
f\left( \tau \right) =\left( 1-\frac{\tau }{t}\right) e^{i\omega _{A}\tau
}\Theta \left( t-\tau \right) ,  \label{B-11}
\end{equation}%
where $\Theta \left( x\right) $ is the Heaviside unit step function, i.e., $%
\Theta \left( x\right) =1$ for $x\geq 0$, and $\Theta \left( x\right) =0$
for $x<0$. Taking the limit $t\rightarrow +\infty $, the modulation spectrum
$f\left( \omega \right) \rightarrow \delta \left( \omega -\omega _{A}\right)
$, then we obtain the golden rule value%
\begin{equation}
R=2\pi G\left( \omega _{A}\right) .  \label{B-12}
\end{equation}%
The modal profile affects on the decay rate via location of the atom. If the
atom is located at $x_{0}=a/2$ and $y_{0}=b/2$, no photons are radiated into
the TM$_{mn}$ guiding mode with even integer $m$ or $n$ since the guiding
mode are standing waves in the transverse direction. The cutoff frequencies
also affect the decay rate via the local density of states. In Fig.~\ref%
{fig:1}b, we have given a schematic diagram of the dispersion relation of
the guiding modes which interact with the atom at $\vec{r}=\left(
a/2,b/2,z_{0}\right) $ for $a=2b$. If the transition frequency $\omega
_{A}<\Omega _{11}$, SE cannot occur since $\rho \left( \omega \right) =0$.
Since $\rho \left( \omega \right) $ tends to infinite when $\omega
_{A}\rightarrow \Omega _{mn}$, the excited state population decays very
rapidly. When $\omega _{A}$ is located in the frequency band between $\Omega
_{11}$ and $\Omega _{31}$, the TM$_{11}$ guiding mode contribute to the
spontaneous rate. However, there is an enhancement or inhibition of
spontaneous decay depending on the factor $\cos \left( 2\pi z_{0}/\lambda
_{1A}\right) $, where the wave length
\begin{equation}
\lambda _{1A}=\frac{2\pi c}{\sqrt{\omega _{A}^{2}-\Omega _{11}^{2}}}.
\label{B-13}
\end{equation}%
In Fig.~\ref{fig:2}(a), we have plotted the probability of finding the atom
in its excited state as a function of $\Gamma t$ for three different values
of $z_{0}=0,\lambda _{1A}/8,\lambda _{1A}/4$, where
\begin{equation}
\Gamma =\frac{4d^{2}\Omega _{11}^{2}}{A\epsilon _{0}c\sqrt{\omega
_{A}^{2}-\Omega _{11}^{2}}}.  \label{B-14}
\end{equation}%
It can be seen that in the interval $z_{0}\in \left[ n\lambda _{1A},n\lambda
_{1A}+\lambda _{1A}/4\right] $ with integer $n$, the SE rate decreases as
the atom-mirror separation increases. However in the interval $z_{0}\in %
\left[ n\lambda _{1A}+\lambda _{1A}/4,n\lambda _{1A}+\lambda _{1A}/2\right] $%
, its SE rate increases as $z_{0}$ increases. Since $\cos x$ is a periodical
function of the argument $x$, the figure is only plotted in $z_{0}\in \left[
0,\lambda _{1A}/4\right] $. It can be seen that at $z_{0}=n\lambda
_{1A}+\lambda _{1A}/4$, the SE is completely suppressed. Since we have set
that $a=2b$, TM$_{51}$ and TM$_{13}$ is the third and fourth guiding modes
which might interacting with the atom. When $\Omega _{31}<\omega _{A}<\Omega
_{51}$, the atom interacts with the continua of two guiding modes TM$_{11}$
and TM$_{31}$. The spontaneous rate increases although it still depends on $%
z_{0}$. As $\omega _{A}$ increases, more and more guiding modes are included
to increase the spontaneous rate.
In Fig.~\ref{fig:2}(b), we have plotted the atomic excitation probability
with $z_{0}=\lambda _{1A}/4$ and $\omega _{A}\approx \left( \Omega
_{31}+\Omega _{51}\right) /2$. In this case, the TM$_{11}$ mode does not
contribut to the SE (see the red dashed line), however, the SE is still
enhanced, this enhancement is due to the atomic coupling to the continuum of
TM$_{31}$ mode (see the blue solid line). One can understand this from Eq.(%
\ref{B-12}) that the SE rate of the atom caused by the TM$_{31}$ mode is
also dependent of the factor $\cos \left( 2\pi z_{0}/\lambda _{2A}\right) $,
where $\lambda _{2A}=2\pi c/\sqrt{\omega _{A}^{2}-\Omega _{31}^{2}}$. Since
the wavelength of the radiation emitted by the atom into the continuum is
different for different guiding modes, the SE is generally increased when
more guiding modes interact with the atom.

One can also obtain the spontaneous rate in Eq.(\ref{B-12}) by replacing $%
\tilde{\varepsilon}\left( \tau \right) $ with $\tilde{\varepsilon}\left(
t\right) $ in Eq.(\ref{B-4}) and making the up limit of integral tend to
infinite. Hence, the Markovian approximation yields the same phenomenon in
the context below Eq.(\ref{B-12}), which means that retardation effect is
neglected.


\section{\label{Sec:5}The atomic population of the excited state}


An excited atom relaxes to its ground state accompanied by an release of a
photon to the EM vacuum. In this hollow waveguide, the emitted photon
propagates along the positive and negative $z$ directions. Since the
termination of the waveguide imposes a hard-wall boundary condition on the
field which behaves as a perfect mirror, the photon traveling along the
negative $z$ axis is retroreflected to the atom, and re-excites the atom,
which leads to a non-Markovian type dynamics of the system.

In this section, we study the spontaneous emission dynamics involving the
retardation time for atom located at $\vec{r}=\left( a/2,b/2,z_{0}\right)$
with $a=2b$. For the convenience of discussion, we denote the transversally
confined propagating modes which couple to the atom as TM$_{j}$ with $%
j=\left( m,n\right) $ according to the ascending order of the cutoff
frequencies. By assuming that the transition frequency $\omega _{A}$ is far
away from the cutoff frequencies $\Omega _{j}$., we can expand frequency $%
\omega _{jk}$ around the transition frequency $\omega _{A}$ up to the linear
term
\begin{equation}
\omega _{jk}=\omega _{A}+v_{j}\left( k-k_{j0}\right) ,  \label{C-1}
\end{equation}%
where $k_{j0}=\sqrt{\omega _{A}^{2}-\Omega _{j}^{2}}/c$ is determined by $%
\omega _{jk_{0}}=\omega _{A}$ and the group velocity%
\begin{equation}
v_{j}\equiv \frac{d\omega _{jk}}{dk}|_{k=k_{j0}}=\frac{c\sqrt{\omega
_{A}^{2}-\Omega _{j}^{2}}}{\omega _{A}}  \label{C-2}
\end{equation}%
is different for different TM$_{j}$ guiding modes. We substitute submit Eq.(%
\ref{C-1}) into integro-differential equation (\ref{B-4}). Integrating over
all wave vectors $k$ gives rise to a linear combination of $\delta \left(
t\pm \tau _{j}-\tau \right) $ and $\delta \left( t-\tau \right) $, where $%
\tau _{j}=2z_{0}/v_{j}$ is the time that the emitted photon take the round
trip between the atom and the mirror in the give transverse mode $j$. We
approximately obtain a delay-differential equation
\begin{equation}
\partial _{t}\tilde{\varepsilon}\left( t\right) =-\sum_{j}\Gamma _{j}\left[
\tilde{\varepsilon}\left( t\right) +e^{i\varphi _{j}}\tilde{\varepsilon}%
\left( t-\tau _{j}\right) \Theta \left( t-\tau _{j}\right) \right]
\label{C-3}
\end{equation}%
for the probability amplitude that the atom at time $t$ is in the excited
state, where
\begin{subequations}
\label{C-4}
\begin{eqnarray}
\varphi _{j} &=&2k_{j0}z_{0}=\sqrt{\omega _{A}^{2}-\Omega _{j}^{2}}\frac{%
2z_{0}}{c} \\
\Gamma _{j} &=&\frac{4d^{2}\Omega _{j}^{2}}{A\epsilon _{0}\omega _{A}v_{j}}%
\sin ^{2}\frac{m\pi }{2}\sin ^{2}\frac{n\pi }{2}.
\end{eqnarray}%
The first term on the right hand side of Eq.(\ref{C-3}) leads to the
exponential decay of the atom. The second term involved the time $\tau _{j}$
that the light needs for the distance atom-mirror-atom, which represents the
effect of the reflected radiation on the atom that was emitted at time $\tau
_{j}$ in the TM$_{mn}$ mode before it interacts again with the atom.

\subsection{SE dynamics in single mode}

In the frequency band between $\Omega _{11}$ and $\Omega _{31}$, the
waveguide is said to be single-moded. The atom with the transition frequency
$\omega _{A}\in \left( \Omega _{11},\Omega _{31}\right) $ only interacts
with the TM$_{11}$ ($j=1$) guiding mode, the delay-differential equation
reduces to
\end{subequations}
\begin{equation}
\partial _{t}\tilde{\varepsilon}\left( t\right) =-\Gamma _{1}\left[ \tilde{%
\varepsilon}\left( t\right) +e^{i\varphi _{1}}\tilde{\varepsilon}\left(
t-\tau _{1}\right) \Theta \left( t-\tau _{1}\right) \right] .  \label{C1-1}
\end{equation}%
where $\Gamma _{1}=\Gamma $ given in Eq. (\ref{B-14}). For the case that the
retarded argument $\tau _{1}\rightarrow 0$, the memory effects inherent in
the system disappear. The amplitude of state $\left\vert e\right\rangle $
becomes
\begin{equation*}
\tilde{\varepsilon}\left( t\right) =\exp \left[ -\Gamma \left( 1+e^{i\varphi
_{1}}\right) t\right] .
\end{equation*}%
The SE rate and the frequency shift are presented by the real part $2\Gamma
\left( 1+\cos \varphi _{1}\right) $ and imagine part $\Gamma \sin \varphi
_{1}$, respectively. In the limit $\tau _{1}\rightarrow \infty $, the second
term of Eq.(\ref{C1-1}) vanishes. Since the waveguide becomes infinite, the
atomic population decays exponentially and the SE rate $\Gamma $ is
independent of the coordinate $z_{0}$.
\begin{figure}[tbp]
\includegraphics[width=9 cm]{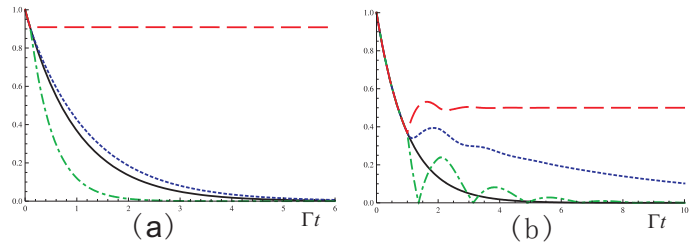}
\caption{\textit{(Color on line)} Amplitude $|\tilde{\protect\varepsilon}%
\left( t\right) |$ versus $\Gamma t$ with delay $\Gamma \protect\tau %
_{1}=0.1 $ (a) and $\Gamma \protect\tau _{1}=1$ (b) in the following cases:
no termination (black solid line), phase $\protect\varphi _{1}=2n\protect\pi %
+\protect\pi /2$ (blue dotted line), $\protect\varphi _{1}=2n\protect\pi +%
\protect\pi $ (red dashed line), $\protect\varphi _{1}=2n\protect\pi $
(green dash-dotted line). We have set the following parameters: $a=2b$, $%
\protect\omega _{A}=(\Omega _{11}+\Omega _{31})/2$. Since both $\protect%
\varphi _{1}$ and $\protect\tau _{1}$ are related to the distance $z_{0}$,
different phases are achieved by adjusting the ratio $a/d$. }
\label{fig:3}
\end{figure}

It can be seen from Eq.(\ref{C1-1}) that the time axis is divided into
intervals of length $\tau _{1}$. We can formally integrate Eq.(\ref{C1-1})
and change the dummy integration variable, which is then substitute into the
integrand. Proceeding indefinitely with iteration, the time behavior of the
atomic state populations reads%
\begin{equation}
\tilde{\varepsilon}\left( t\right) =\sum_{l=0}^{\infty }\frac{\left( -\Gamma
e^{i\varphi _{1}}\right) ^{n}}{n!}e^{-\Gamma \left( t-l\tau _{1}\right)
}\left( t-l\tau _{1}\right) ^{n}\Theta \left( t-l\tau _{1}\right) .
\label{C1-2}
\end{equation}%
A step character is presented in Eq.(\ref{C1-2}). For $t\in \left[ 0,\tau
_{1}\right] $, the atomic amplitude $\tilde{\varepsilon}\left( t\right)
=\exp \left( -\Gamma t\right) $ decays exponentially which coincides with
the behavior of a excited atom in an infinite waveguide. The underlying
physics is that the atom requires at least the time $\tau _{1}$ to recognize
the mirror. For $t\in \left[ \tau _{1},2\tau _{1}\right] $, due to the
emitted radiation reflected back to the atom, $l=1$ term has been included,
which gives rise to the interference for finding the atom in the excited
state. In Fig.~\ref{fig:3}, we have plotted the norm $\left\vert \tilde{%
\varepsilon}\left( t\right) \right\vert $ of the atomic amplitude versus $%
\Gamma t$ with delay $\Gamma \tau _{1}=0.1$ (a) and $\Gamma \tau _{1}=1$
(b). The exponential decay of the atom inside an infinite waveguide is
plotted with the black solid line. When $\Gamma \tau _{1}\ll 1$, an
exponential decay law is found, however, the SE can be either increased or
descreased by the phase. The SE is completely suppressed when phase $\varphi
_{1}=\left( 2n+1\right) \pi $. When $\Gamma \tau _{1}\geq 1$, the atom first
decays exponentially, after the atom recognize the mirror, it displays a
behavior deviating from the exponential decay law, and a partial revival of
the atomic population can be found. It it the interference between the
emitted wave and the radiation wave reflected back to the atom which makes
the atom-mirror separation $z_{0}$ has significant influence on the atomic
dynamic via the phase. When the distance between the atom and the
termination are further large (i.e., $\Gamma \tau _{1}\gg 1$), it is found
from Fig.~\ref{fig:5}a that there is also a partial revival of the atomic
population, however, the atom-mirror separation $z_{0}$ does not make any
sense. In this case, the atom have already decayed to the ground state at
the time that the photon bounces back to the atom, so there is no emitted
wave to be interference with the wave reflected back to the atom, which
means that the atomic revival is due to the atom being partially re-excited
by the radiation. Since the the light emitted in the positive direction has
depart from the atom, the probability that the atom is re-excited becomes
lower and lower.

\subsection{SE dynamics in multiple modes}

An excited atom radiates waves into the continua of all modes resonant with
the atom. When the cross area become larger, more modes are included in the
resonance, then the atomic dynamics is not only affected by the time $\tau
_{1}$ that light needs to bounce back and forth between the atom and the
termination in the TM$_{11}$ mode, but also by other time $\tau _{j}$
required for a photon emitted by the atom to propagate in the TM$_{mn}$ mode
and reabsorbed by the atom. The definition of delay time $\tau _{j}$ told us
that $\tau _{j}<\tau _{j+1}$.
\begin{figure}[tbp]
\includegraphics[width=9 cm]{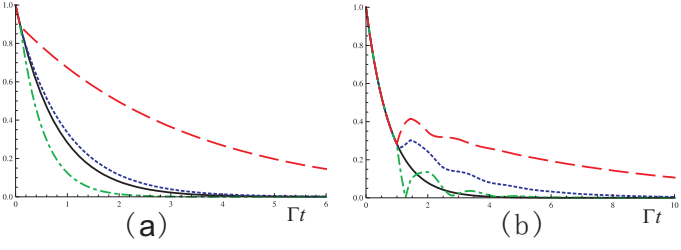}
\caption{\textit{(Color on line)} Amplitude $|\tilde{\protect\varepsilon}%
\left( t\right) |$ versus $\Gamma t$ with delay $\Gamma \protect\tau %
_{1}=0.1 $ (a) and $\Gamma \protect\tau _{1}=1$ (b) in the following cases:
no termination (black solid line), phase $\protect\varphi _{1}=2n\protect\pi %
+\protect\pi /2$ (blue dotted line), $\protect\varphi _{1}=2n\protect\pi +%
\protect\pi $ (red dashed line), $\protect\varphi _{1}=2n\protect\pi $
(green dash-dotted line). We have set the following parameters: $a=2b$, $%
\protect\omega _{A}=(\Omega _{31}+\Omega _{51})/2$. }
\label{fig:4}
\end{figure}

In this section, we assume that the atomic transition frequency is in the
regime $\left[ \Omega _{31},\Omega _{51}\right] $, which means that only two
TM modes (i.e. TM$_{11}$, TM$_{31}$) in resonance with the atom, the
delay-differential equation reduces to%
\begin{eqnarray}
\partial _{t}\tilde{\varepsilon}\left( t\right) &=&-\left( \Gamma +\Gamma
_{2}\right) \tilde{\varepsilon}\left( t\right) -\Gamma e^{i\varphi _{1}}%
\tilde{\varepsilon}\left( t-\tau _{1}\right) \Theta \left( t-\tau _{1}\right)
\notag  \label{C2-1} \\
&&-\Gamma _{2}e^{i\varphi _{2}}\tilde{\varepsilon}\left( t-\tau _{2}\right)
\Theta \left( t-\tau _{2}\right) .
\end{eqnarray}%
The space dependence enters via both the phases $\varphi _{1}$, $\varphi
_{2} $ and the delay time $\tau _{1},\tau _{2}$ of the different modes. If
both arguments $\tau _{1},\tau _{2}\rightarrow 0$, the amplitude of state $%
\left\vert e\right\rangle $ becomes%
\begin{equation}
\tilde{\varepsilon}\left( t\right) =\exp \left[ -\sum_{j=1}^{2}\Gamma
_{j}\left( 1+e^{i\varphi _{j}}\right) t\right]  \label{C2-2}
\end{equation}%
Two terms are consisted of in the above equation, the SE rate $%
\sum_{j=1}^{2}2\Gamma _{j}\left( 1+\cos \varphi _{j}\right) $ and frequency
shift $\sum_{j=1}^{2}\Gamma _{j}\sin \varphi _{j}$. Comparing to the single
mode case, the SE rate is enhanced, however the frequency shift can be
either increased or decreased due to the space dependence. In the limit $%
\tau _{1},\tau _{2}\rightarrow \infty $, the second and third terms of Eq.(%
\ref{C2-1}) vanishes. The amplitude $\tilde{\varepsilon}\left( t\right)
=\exp \left[ -\left( \Gamma +\Gamma _{2}\right) t\right] $ shows that the
atomic population decays exponentially, $\Gamma +\Gamma _{2}$ is the SE rate
that the atom interacts with the continuum of the TM$_{11}$ and TM$_{31}$
modes of an infinite waveguide, which is also independent of the coordinate $%
z_{0}$. In the case that $\tau _{1}\rightarrow 0$, the delay-differential
equation reads%
\begin{eqnarray}
\partial _{t}\tilde{\varepsilon}\left( t\right) &=&-\left( \Gamma +\Gamma
e^{i\varphi _{1}}+\Gamma _{2}\right) \tilde{\varepsilon}\left( t\right)
\label{C2-3} \\
&&-\Gamma _{2}e^{i\varphi _{2}}\tilde{\varepsilon}\left( t-\tau _{2}\right)
\Theta \left( t-\tau _{2}\right)  \notag
\end{eqnarray}%
Using Laplace transformation and geometric series expansion, the solution
read%
\begin{equation}
\tilde{\varepsilon}\left( t\right) =\sum_{l=0}^{\infty }\frac{\left( -\Gamma
_{2}e^{i\varphi _{2}}\right) ^{l}}{n!}e^{-\left( \Gamma +\Gamma e^{i\varphi
_{1}}+\Gamma _{2}\right) \left( t-l\tau _{2}\right) }\left( t-l\tau
_{2}\right) ^{l}  \label{C2-4}
\end{equation}%
If the atom is located at $z_{0}$ satisfying $\varphi _{1}=\left(
2n+1\right) \pi $, the SE that the TM$_{11}$ mode contribute to is
completely suppressed, then one obtain the SE dynamics due to the emitted
photon propagating only via the continuum of the TM$_{31}$ mode. In the case
with finite $\tau _{1}$ and $\tau _{2}\rightarrow \infty $, the upper state
amplitude becomes%
\begin{equation}
\partial _{t}\tilde{\varepsilon}\left( t\right) =-\left( \Gamma +\Gamma
_{2}\right) \tilde{\varepsilon}\left( t\right) -\Gamma e^{i\varphi _{1}}%
\tilde{\varepsilon}\left( t-\tau _{1}\right) \Theta \left( t-\tau
_{1}\right) .  \label{C2-5}
\end{equation}%
However, it is impossible for an atom inside a realistic waveguide to appear
the dynamics described by Eq.(\ref{C2-5}).

In Fig.~\ref{fig:4}, we have numerically plot the amplitude $\left\vert
\tilde{\varepsilon}\left( t\right) \right\vert $ as a function of $\Gamma t$
with $\Gamma \tau _{1}=0.1$ (a) and $1$ (b). It can be seen that in the
interval $\left[ 0,\tau _{1}\right] $, the upper state population of the
atom decays exponential with a rate $\Gamma +\Gamma _{2}$. After time $\tau
_{1}$, photons emitted by the atom is reflected back to the atom by the
mirror so that the atom-mirror distance has great influence on the SE
dynamics via phase $\varphi _{j}$ and $\tau _{j}$. Phase $\varphi _{1}$
first gives arise to deviation from the decay with a rate $\Gamma +\Gamma
_{2}$ in the interval $\left[ \tau _{1},\tau _{2}\right] $. As soon as $%
t>\tau _{2}$, the wave propagating in the TM$_{31}$ mode is reflected back
to the atom by the mirror, phase $\varphi _{2}$ deviates the atomic dynamics
from that of finite $\tau _{1}$ and $\tau _{2}\rightarrow \infty $. In the
weak coupling case (see, Fig.~\ref{fig:4}a), the excited state probability
decreases as the time increases. However, in the strong coupling case,
several peaks can be observed in Fig.~\ref{fig:4}b, which present partial
revivals of the atom probability when $\Gamma \tau _{1}\geq 1$. Comparing to
the time evolution in Fig.~\ref{fig:3}, the SE is enhanced for a given phase
$\varphi _{1}$. Phase $\varphi _{2}$ and retardation time $\tau _{2}$ shift
the position of the peak and the dip.
\begin{figure}[tbp]
\includegraphics[width=9 cm]{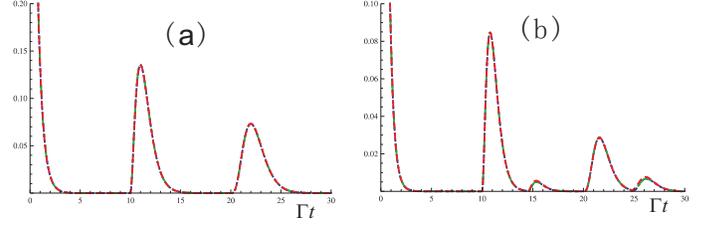}
\caption{\textit{(Color on line)} Probability $|\tilde{\protect\varepsilon}%
\left( t\right) |^2$ versus $\Gamma t$ with delay $\Gamma\protect\tau %
_{1}=10 $ with phase $\protect\varphi _{1}=2n\protect\pi +\protect\pi /2$
(blue dotted line), $\protect\varphi _{1}=2n\protect\pi +\protect\pi $ (red
dashed line), $\protect\varphi _{1}=2n\protect\pi $ (green dash-dotted line)
in the following cases: (a) single mode $\protect\omega _{A}=(\Omega _{31}
+\Omega _{11})/2$, (b) two modes $\protect\omega _{A}=(\Omega _{31}+\Omega
_{51})/2$. Here, $a=2b$.}
\label{fig:5}
\end{figure}

In Fig.~\ref{fig:5}, we have numerically plotted the probability $|\tilde{%
\varepsilon}\left( t\right)|^2$ as a function of $\Gamma t$ with delay $%
\Gamma\tau _{1}=10$ for (a) single-moded and (b) double-moded cases. When
the photon returns back to the atom, the atom has already decay to its
ground state, it is impossible for interference to occur so that the phases $%
\varphi_j$ has no effect on the atomic dynamics. The peaks at time $t>\tau
_{1}$ ow to the reflected light reabsorbed by the atom. By comparing two
figures in Fig.~\ref{fig:5}, we found that the probability that the atom is
reexcited by the radiation wave is lower in the multiple-mode case than that
in the single-mode case, and more peaks appear in the interval $[m\tau
_{1},(m+1)\tau _{1}]$ for the multiple-mode case. Such observations are easy
to understand because there are more transverse modes to interact with the
atom.


\section{discussion and conclusion}


We have study the dynamics of an atom inside a hollow waveguide of
rectangular cross section $A=ab$, made of ideal perfect conducting walls.
Such 1D waveguide generally consists of both TE and TM waves, the atom with
dipole along the $z$-direction interacts only with the TM$_{mn}$ transverse
modes, their coupling strength depends on the atomic location. A two-level
atom with location fixed at $\left( a/2,b/2,z_{0}\right) \ $is considered,
which decouples to fields of the TM$\ $modes with even integer $m,n$. We
have first discussed the dependence of the SE rate on the atom-mirror
separation and the density of states by Markovian approximation. 1) Since
the density of state vanishes below $\Omega _{11}$, the SE is completely
suppressed when $\omega _{A}<\Omega _{11}$; 2) Since the density of state
tends to infinite, the excited state population decays very rapidly when $%
\omega _{A}\rightarrow \Omega _{mn}$, 3) Away from the cutoff frequencies,
the SE rate is increased when more TM modes in resonance with the atom, and
the upper state population could be unchanged all the time for single mode
case with $z_{0}=n\lambda _{1A}+\lambda _{1A}/4$ (i.e., $\varphi _{1}=\left(
2n+1\right) \pi $). However, the distance $z_{0}\neq 0$ gives rise to the
time delay that radiation emitted by the atom return to the emitter. To
study this backaction of the ideal perfect conducting at $z=0$ on the atom,
we perform a linear approximation, which is valid for $\omega _{A}$ far from
the cutoff frequencies, and phase $\varphi _{j}$ and retardation time $\tau
_{j}$ are introduced into the dynamics of the atom via $\omega _{A}$ and $%
z_{0}$. The SE dynamics is studied for both single-moded and double-moded
cases. We find that 1) the upper state population is less than its initial
as long as $\tau _{j}\neq 0$. The discussion in Markov approximation
corresponds to the case with the retardation time $\tau _{j}\rightarrow 0$.
2) The probability for finding the atom in its excited state is lowered when
more transverse modes are resonant with the atom. 3) In the interval $\left[
0,\tau _{1}\right] $, the atomic behavior is the same to that of an excited
atom in an infinite waveguide, and the SE rate is independent of $z_{0}$. 4)
After $t>\tau _{1}$, two situations should be distinguished. For short
retardation time, the interference between the radiation wave and the
emitted wave makes the dynamics is strongly dependent on $\varphi _{j}$. For
long retardation time, the atom has already decay to its ground state when
the photon returns back to the atom, the partial revivals and collapses are
due to the photon reabsorbed and re-emitted by the atom. .

\begin{acknowledgments}
This work was supported by NSFC Grants No. 11374095, No. 11422540, No.
11434011, No. 11575058; National Fundamental Research Program of China (the
973 Program) Grant No. 2012CB922103; Hunan Provincial Natural Science
Foundation of China Grants No. 11JJ7001.
\end{acknowledgments}

\end{document}